\documentclass[fleqn,12pt,twoside]{article}
\usepackage{espcrc1}

\readRCS $Id: espcrc1.tex,v 1.2 2004/02/24 11:22:11 spepping Exp $
\usepackage{graphicx}
\usepackage[figuresright]{rotating}

\newcommand{\AmS}{{\protect\the\textfont2
  A\kern-.1667em\lower.5ex\hbox{M}\kern-.125emS}}

\title{The transverse energy per charged particle estimates in the
framework of a statistical model}

\author{Dariusz Prorok\address[MCSD]{Institute of Theoretical Physics, University of
Wroc{\l}aw, \\
        Pl.Maksa Borna 9, 50-204  Wroc{\l}aw, Poland}%
        \thanks{Supported in part by the Polish State Committee for Scientific
        Research, grant KBN 2 P03B 069 25.}}

\begin{document}

\maketitle

\begin{abstract}
The transverse energies and the charged particle multiplicities at
midrapidity as well as their ratio, $dE_{T}/d\eta\mid_{mid} /
dN_{ch}/d\eta\vert_{mid}$, are evaluated in a statistical model
with expansion for the wide range of heavy-ion collisions, from
AGS to RHIC at $\sqrt{s_{NN}}=200$ GeV. Full description of decays
of hadron resonances is applied in calculations of both
$dE_{T}/d\eta\vert_{mid}$ and $dN_{ch}/d\eta\vert_{mid}$. The
predictions of the model at the freeze-out parameters, established
independently from observed particle yields and $p_{T}$ spectra,
agree well with the experimental data.
\end{abstract}

\section*{}

The statistical model has succeeded in the description of particle
yield ratios and $p_{T}$ spectra measured in heavy-ion collisions
\cite{Braun-Munzinger:1994xr,Braun-Munzinger:1995bp,Cleymans:1996cd,Stachel:wh,Braun-Munzinger:1999qy,Becattini:2000jw,Braun-Munzinger:2001ip,Florkowski:2001fp,Broniowski:2001we,Broniowski:2001uk,Baran:2003nm,Broniowski:2002nf,Michalec:2001um,Broniowski:2002am}.
Transverse energy ($dE_{T}/d\eta$) and charged particle
multiplicity densities ($dN_{ch}/d\eta$) are global observables
whose measurements are independent of hadron spectroscopy,
therefore they could be used as an additional test of the
self-consistency of the statistical model.

The experimentally measured transverse energy is defined as

\begin{equation}
E_{T} = \sum_{i = 1}^{L} \hat{E}_{i} \cdot \sin{\theta_{i}} \;,
\label{Etdef}
\end{equation}

\noindent where $\theta_{i}$ is the polar angle, $\hat{E}_{i}$
denotes $E_{i}-m_{N}$ ($m_{N}$ means the nucleon mass) for baryons
and the total energy $E_{i}$ for all other particles, and the sum
is taken over all $L$ emitted particles \cite{Adcox:2001ry}.
Additionally, in the case of RHIC at $\sqrt{s_{NN}}=200$ GeV,
$E_{i}+m_{N}$ is taken instead of $E_{i}$ for antibaryons
\cite{Bazilevsky:2002fz}.
\begin{table}[htb]
\caption{Values of $dE_{T}/d\eta\vert_{mid}$ and
$dN_{ch}/d\eta\vert_{mid}$ calculated in the framework of the
statistical model with expansion. In the first column thermal and
geometric parameters are listed for the corresponding collisions.
The data are for the most central collisions.} \label{Table1}
\begin{tabular}{@{}lcccc}
\hline Collision case &
\multicolumn{2}{c}{$dE_{T}/d\eta\vert_{mid}$ [GeV]} &
\multicolumn{2}{c}{$dN_{ch}/d\eta\vert_{mid}$}
\\
\cline{2-3}\cline{4-5} & Theory & Experiment & Theory & Experiment
\\
\hline Au-Au at RHIC, $\sqrt{s_{NN}}=200$ GeV: & & & &
\\
 $T
= 165.6$ MeV, $\mu_{B} = 28.5$ MeV & $585^{\mathrm{(a)}}$ & $597
\pm 34$ \protect\cite{Bazilevsky:2002fz} & 589 & $699 \pm 46$
\protect\cite{Bazilevsky:2002fz}
\\
 $\rho_{max} = 7.15$ fm, $\tau
= 7.86$ fm \protect\cite{Baran:2003nm} & & & & $579 \pm
29^{\mathrm{(b)}}$
\\
 & & & & \protect\cite{Adler:2003cb} \\  Au-Au at
RHIC, $\sqrt{s_{NN}}=130$ GeV: & & & &
\\
 $T = 165$ MeV,
$\mu_{B} = 41$ MeV & 507 & $503 \pm 25$
\protect\cite{Adcox:2001ry} & 555 & $622 \pm 41$
\protect\cite{Adcox:2000sp}
\\
 $\rho_{max} = 6.9$ fm, $\tau =
8.2$ fm \protect\cite{Broniowski:2002nf} & & & & $568 \pm
47^{\mathrm{(b)}}$
\\
 & & & & \protect\cite{Adcox:2001mf}
\\
 Pb-Pb
at SPS: & & & & \\
 $T = 164$ MeV, $\mu_{B} = 234$ MeV & 447 &
$363 \pm 91$ \protect\cite{Aggarwal:2000bc} & 476 &
$464_{-13}^{+20}$ \protect\cite{Aggarwal:2000bc}
\\ $\rho_{max} =
6.45$ fm, $\tau = 5.74$ fm
\protect\cite{Michalec:2001um,Broniowski:2002am} & & & & \\
 & & & & \\ Au-Au at AGS: & & & &
\\ $T = 130$ MeV, $\mu_{B} = 540$ MeV & 224 & $\approx 200$
\protect\cite{Barrette:pm} & 271
& $\approx 270$ \protect\cite{Barrette:1994kr} \\
$\beta_{\perp}^{max} = 0.675$, $\rho_{max} = 6.52$ fm
\protect\cite{Braun-Munzinger:1994xr,Stachel:wh} & & & & \\ & & &
& \\  Si-Pb at AGS: & & & & \\  $T = 120$ MeV, $\mu_{B} = 540$ MeV
& 57 & $\approx 62$ \protect\cite{Barrette:1994kr} & 91 & $\approx
115-120$ \\ $\beta_{\perp}^{max} = 0.54$, $\rho_{max} = 5.02$ fm
\protect\cite{Braun-Munzinger:1994xr,Stachel:wh} & & & &
\protect\cite{Barrette:1994kr}
\\
\hline
\end{tabular}\\[2pt]
$^{\mathrm{(a)}}$ \scriptsize{For the modified definition of
$E_{T}$, \emph{i.e.} $E_{i}+m_{N}$ is taken instead of $E_{i}$ for
antibaryons, see eq.~(\protect\ref{Etdef}).}
\par $^{\mathrm{(b)}}$ \scriptsize{For the charged particle
multiplicity expressed as the sum of integrated charged hadron
yields.}
\end{table}
\begin{table}[htb]
\caption{Values of the ratio $dE_{T}/d\eta\vert_{mid}/
dN_{ch}/d\eta\vert_{mid}$ calculated in the framework of the
statistical model with expansion. The data are for the most
central collisions.} \label{Table2}
\begin{center}
\begin{tabular}{@{}ccc} \hline Collision case & \multicolumn{2}{c}{
$dE_{T}/d\eta\vert_{mid}/ dN_{ch}/d\eta\vert_{mid}$ [GeV]} \\
\cline{2-3} & Theory & Experiment \\
\hline Au-Au at RHIC at $\sqrt{s_{NN}}=200$ GeV &
$0.99^{\mathrm{(a)}}$ & $0.87 \pm 0.06$
\protect\cite{Bazilevsky:2002fz} \\ & & $1.03 \pm
0.08^{\mathrm{(b)}}$
\\
Au-Au at RHIC at $\sqrt{s_{NN}}=130$ GeV & 0.91 & $0.81 \pm 0.06$
\protect\cite{Adcox:2001ry} \\ & & $0.89 \pm 0.09^{\mathrm{(b)}}$
\\ Pb-Pb at SPS
& 0.94 & $0.78 \pm 0.21$ \protect\cite{Aggarwal:2000bc} \\ & & \\
Au-Au
at AGS & 0.83 & $0.72 \pm 0.08$ \protect\cite{Barrette:1994kr} \\
& & \\ Si-Pb at AGS & 0.63 & 0.52-0.54
\protect\cite{Barrette:1994kr} \\ \hline
\end{tabular}
\end{center}
\par \hspace{1cm} $^{\mathrm{(a)}}$ \scriptsize{For the modified definition of
$E_{T}$, \emph{i.e.} $E_{i}+m_{N}$ is taken instead of $E_{i}$ for
antibaryons, see eq.~(\protect\ref{Etdef}).}
\par \hspace{1cm} $^{\mathrm{(b)}}$ \scriptsize{Author calculations with the use
of experimental values given in table\,\protect\ref{Table1} and
the denominator \par \hspace{1.5cm} expressed as the sum of
integrated charged hadron yields.}
\end{table}

The statistical model with single freeze-out
\cite{Broniowski:2001we,Broniowski:2001uk,Broniowski:2002nf} is
applied for evaluations of $dE_{T}/d\eta$ and $dN_{ch}/d\eta$ at
midrapidity. Details of this analysis can be found elsewhere
\cite{Prorok:2002ta,Prorok:2004af}. The foundations of the model
are as follows: (\textit{a}) the chemical and thermal freeze-outs
take place simultaneously, (\textit{b}) all confirmed resonances
up to a mass of $2$ GeV from the Particle Data Tables
\cite{Hagiwara:fs} are taken into account, (\textit{c}) a
freeze-out hypersurface is defined by the equation $\tau =
\sqrt{t^{2}-r_{x}^{2}-r_{y}^{2}-r_{z}^{2}}= const$, (\textit{d})
the four-velocity of an element of the freeze-out hypersurface is
proportional to its coordinate, $u^{\mu}= x^{\mu} / \tau$,
(\textit{e}) the transverse size is restricted by the condition
$r=\sqrt{r_{x}^{2}+r_{y}^{2}}< \rho_{max}$. The maximum
transverse-flow parameter is expressed as
$\beta_{\perp}^{max}=(\rho_{max}/\tau)
/(\sqrt{1+(\rho_{max}/\tau)^{2}})$. The model has four parameters,
namely, the two thermal parameters, the temperature $T$ and the
baryon number chemical potential $\mu_{B}$, and the two geometric
parameters, $\tau$ and $\rho_{max}$. Values of these parameters
were established from fits to particle yield ratios and $p_{T}$
spectra (see the first column of table\,\ref{Table1}). The
invariant distribution of the measured particles of species $i$
has the Cooper-Frye form
\cite{Broniowski:2001we,Broniowski:2001uk}. The distribution
collects, besides the thermal one, also contributions from simple
and sequential decays such that at least one of the final
secondaries is of the \emph{i} kind (for details, see
\cite{Broniowski:2002nf,Prorok:2004af}). Having integrated this
distribution suitably over $p_{T}$ and summing up over final
particles, one can obtain $dE_{T}/d\eta$ and $dN_{ch}/d\eta$. The
results together with the corresponding experimental data are
listed in table\,\ref{Table1}.

\begin{figure}[htb]
\includegraphics[scale=0.6]{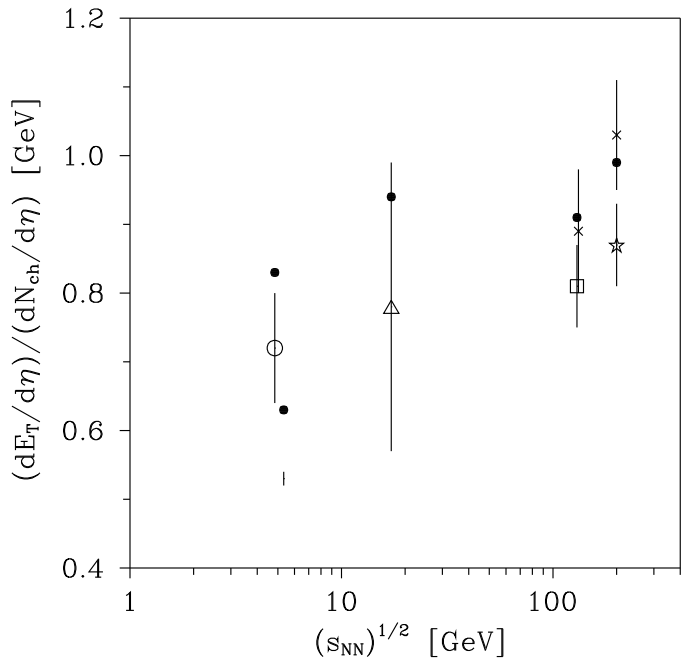}
\caption{Values of the transverse energy per charged particle at
midrapidity for the most central collisions. Black dots denote the
model evaluations. Also data points for AGS
\protect\cite{Barrette:1994kr} (a circle for Au-Au and a vertical
bar for Si-Pb), SPS \protect\cite{Aggarwal:2000bc} (triangle),
RHIC at $\sqrt{s_{NN}}=130$ GeV \protect\cite{Adcox:2001ry}
(square) and RHIC at $\sqrt{s_{NN}}=200$ GeV
\protect\cite{Bazilevsky:2002fz} (star) are depicted. For RHIC,
points with the sum of integrated charged hadron yields
substituted for the denominator are also depicted (crosses).}
\label{Fig.1.}
\end{figure}

Generally, the predictions agree well with the data. However, the
$15\%$ underestimation of the charged particle density has been
found for RHIC. This can be explained by the existing
inconsistency in measurements of the charged particle multiplicity
at RHIC. Namely, the sum of integrated charged hadron yields is
substantially below the directly measured
$dN_{ch}/d\eta\vert_{mid}$ (see table\,\ref{Table1}). Note that
values of this sum agree very well with the model predictions.
Estimates of the ratio $dE_{T}/d\eta\vert_{mid} /
dN_{ch}/d\eta\vert_{mid}$ are collected in table\,\ref{Table2}
together with the corresponding data. These results have been also
depicted in fig.\,\ref{Fig.1.}. The $15\%$ overall overestimation
has been obtained. In the RHIC case the mentioned underestimation
of $dN_{ch}/d\eta\vert_{mid}$ is the reason for that. But if
$dN_{ch}/d\eta\vert_{mid}$ from the summing up of integrated
hadron yields is put in the denominator, the theoretical
predictions agree very well with the data.

To conclude, the thermal model has been used to reproduce
transverse energy and charged particle multiplicity pseudorapidity
densities and their ratio from AGS, SPS and RHIC. The evaluations
have been made at the thermal and geometric freeze-out parameters
obtained in previous analyses of measured particle ratios and
$p_{T}$ spectra. The good accuracy of predictions has confirmed,
in an alternative way, the applicability of the thermal model to
the description of the soft part of the particle production in
heavy-ion collisions.

\end{document}